%% file: seyfert.tex
\documentstyle[aas2pp4,amssym,flushrt]{article}

\begin{document}

\title{Compact Radio Cores and Nuclear Activity in Seyfert Galaxies}
\author{ Giuliano GIURICIN, Dario FADDA, Marino MEZZETTI} 
\affil{
Dipartimento di Astronomia, Universit\`{a} degli Studi di Trieste, \\
SISSA, via Beirut 4, 34013 - Trieste, Italy\\ E-mail:
giuricin@sissa.it; fadda@sissa.it; mezzetti@sissa.it}

\begin{abstract} 
Using recent high--resolution ($<$0.1") radio observations of a large
sample of Seyfert galaxies (Roy et al., 1994), we analyze the
relations between their compact radio core emission and several
nuclear and host galaxy properties (galaxy morphology, optical,
infrared, X-ray, extended radio emissions, interaction parameters, and
some emission line properties). We apply survival analysis techniques
in order to exploit the information contained in the numerous
"censored" data (upper limits on fluxes).

We find that Seyfert nuclei hosted in early--type galaxies are, on
average, characterized by stronger radio core emission than the norm
for Seyfert galaxies.  Galaxies with a nearby companion display
enhanced radio core emission with respect to objects without
companions. Furthermore, we confirm that Seyfert types 2 host more
powerful compact radio cores than types 1.

Remarkably, radio core emission appears to be unrelated to optical,
near--infrared, and far--infrared radiations, but shows some
correlation with total radio emission and with tracers of nuclear
activity such as the IRAS 12 and 25 $\mu$m band, hard X-ray and
narrow--line emissions. This favours the view that Seyfert radio cores
are typically powered by AGN rather than by radio supernovae.

A link between radio core strength and bolometric luminosity is
suggested, in analogy to what is observed in the cores of radio--quiet
QSOs.

\end{abstract} 

\keywords{ galaxies: active -- galaxies: Seyfert -- galaxies: nuclei
-- radio continuum: galaxies }

\section{INTRODUCTION} 

Very high resolution radio observations of active galactic nuclei
(AGN) can be profitably used to probe the central engine, since they
are principally sensitive to nuclear non-thermal emission at high
brightness temperature from AGN activity and help us to discriminate
against low brightness temperature extended radio emission expected
from the presence of circumnuclear HII regions. For instance, compact
radio cores are fairly frequented detected in samples of Seyfert
galaxies, but rarely detected in optically selected starburst galaxies
(e.g., Norris et al., 1990).  Consistently, the deeper, very long
baseline interferometry (VLBI) survey of ultraluminous infrared
(IR) galaxies by Lonsdale, Smith \& Lonsdale (1993), which did not
reveal any correlation between the detection of compact radio cores
and the optical classification of galactic nuclei, has been used to
suggest the likely existence of AGNs embedded in dust within their
nuclei.
The observations of compact radio cores in Seyfert galaxies have
recently been also employed as a new test of the canonical unification
schemes of AGNs, according to which Seyfert type 1 objects (S1) would
contain the same central engines as Seyfert types 2 (S2), but would
have the broad--line region (BLR) and the strong optical, UV, and
X-ray continua of the central source concealed from our view by a disk
or torus of dusty molecular clouds (see, e.g., the review by
Antonucci, 1993). If the torus surrounding the nucleus is optically
thin at radio wavelengths, orientation effects should have no effect
on the radio appearance of Seyfert galaxies. As a matter of fact,
contrarily to old contentions (see, e. g., the review by Lawrence,
1987) based on poor statistics, recent low--resolution radio
observations show no significant differences in the major radio
properties (central and total radio powers, radio size, radio spectral
index) of S1 and S2 objects (e.g., Edelson, 1987; Ulvestad \& Wilson,
1989; Giuricin et al., 1990), as is expected from standard unified
schemes.

But, surprisingly, in the high--resolution radio survey by Roy et al.
(1994) compact radio cores were found more commonly in S2 than in S1
galaxies, although the disk radio powers, [OIII]$\lambda 5007$\AA $\!$
emission line luminosities, and far--infrared (FIR) luminosities of
the two classes of objects are similar (Roy et al., 1994, 1996). In
order to reconcile their findings with the unified view of AGNs, the
authors offered some models which resort to free--free absorption of
S1 radio cores by the narrow--line region (NLR), if the radio cores
lie in the BLR, or by individual NLR clouds, if the radio cores are
located in the NLR.

In this paper we use Roy et al.'s (1994) radio core data set in order
to examine the relations (as yet unexplored) between radio core
emission and several properties of Seyfert galaxies and
nuclei. Specifically, we examine the 2.3 GHz radio core fluxes $F_c$
of the optical+infrared--selected sample of Seyfert galaxies observed
by Roy et al. (1994). The authors observed 157 Seyfert galaxies with
the 275 km Parkes--Tidbinbilla Interferometer (PTI) at 1.7 or 2.3
GHz. They combined fluxes (or 5$\sigma$ upper limits on fluxes) from
different observations, converting 1.7 GHz fluxes to 2.3 GHz with the
adoption of a spectral index $n=-0.7 ~ (F_{\nu}\propto\nu^{n})$. Their
high--resolution survey, characterized by uniform sensitivity and
resolution, is blind to Kpc--scale emission of low brightness
temperature, whereas it is sensitive only to structures with
brightness temperature greater than $10^{5}$ K and sizes less than
0".1, which correspond to 20--200 $pc$ over the redshift range
$0.01<z<0.1$, typical of the sample.

Radio core data of 22 Seyfert galaxies have recently been published by
Sadler et al. (1995) as part of their multifrequency (at 1.7 GHz, 2.3
GHz, 8.4 GHz) PTI survey of radio cores in a sample of nearby spiral
galaxies. Their radio core fluxes appear to be in substantial
agreement with Roy et al.'s (1994) for the 10 objects in common.
 
In this paper we restrict ourselves to Roy et al.'s (1994) 149 Seyfert
galaxies having $z<0.1$, since distant galaxies might represent a
different class of objects. The radio data set considered comprises
about 35\% of detections and 65\% of upper limits (25\% and 47\% of
detections for S1 and S2, respectively). The presence of many upper
limits on the radio fluxes dictates the use of the techniques of
Survival Analysis suitable for the treatment of "censored" data (see,
e. g., the review by Feigelson, 1992). In practice we use the software
package ASURV (Rev 1.2, La Valley, Isobe \& Feigelson, 1992) in which
the methods presented in Feigelson \& Nelson (1985) and in Isobe,
Feigelson \& Nelson (1986) are implemented. We pay particular
attention to removing distance effects when we study
luminosity--luminosity correlations and when we compare luminosities
of subsets of objects which lie at different average distances.

Basically, we search for correlations (as yet unexplored) between the
radio core emission and several properties of Seyfert nuclei and host
galaxies. These include galaxy morphology, optical, infrared, X-ray,
extended radio emissions, interaction parameters and some
emission--line nuclear properties. Several of these properties are
indicators of nuclear activity.  Our study will also ensure that the
above-mentioned difference between the radio core emissions of S1 and
S2 objects is not simply due to a systematic difference of the two
samples in some properties, to which radio core emission might be
strongly related.

In \S 2 we describe our statistical analysis. In some tables or plots
we present the most outstanding results for the interesting cases. We
shall not bother with extremely weak correlations (at the $<$90\%
significance level); correlations at a level falling between 90\% and
95\% will be referred to as marginal. All data compiled and results
obtained are available on request. \S 3 summarizes our results and
contains our conclusions.

\section{Analysis and Results} 

\subsection{The Radio Core Powers of Seyfert Galaxies.} 

The distances D of the nearby galaxies which are included in the
"Nearby Galaxies Catalog" (NBG) by Tully (1988) are taken directly
from the values tabulated therein; these distances are based on
velocities, an adopted Hubble constant $H_{0}= 75 km s^{-1} Mpc^{-1}$
and the Virgocentric retardation model described by Tully \& Shaya
(1984). We use redshift--distances with the same $H_{0}$ and $q_0$=0
to calculate the monochromatic radio core powers $P_c$ (expressed in
W/Hz) of non-NBG galaxies.

We employ the Kaplan--Meier product--limit estimator (which is a part
of the ASURV package) in order to calculate the median and the
distribution function of the radio core power $P_c$. The Kaplan--Meier
estimator provides an efficient, non--parametric reconstruction of
information lost by censoring in the case of randomly censored data
sets. This estimator redistributes the upper limits uniformily along
all the intervals of lower detected values. In our case it can be
applied to radio powers, whose censoring pattern tends to be
randomized by the large distance interval covered by our objects.
Figure 1 shows the histogram of $\log P_c$ for the 149 Seyfert
galaxies. For our Seyfert sample (N=149 objects, with $N_{ul}$=93
upper limits) we obtain a median value of $\log P_c= 20.9$ (W/Hz),
together with the 75th and 25th percentiles of 20.1 (W/Hz) and 22.0
(W/Hz). The histogram has a clear maximum at the lowest detected
powers ($\log P_c\sim$ 20 (W/Hz)) and a plateau around $\log P_c$= 22
(W/Hz). The data are consistent with the presence of compact cores in
all Seyfert galaxies at a level of $\log P_c\lesssim$20 (W/Hz), which
falls below the detection threshold for the nearest objects.

\begin{figure}[h]
\plotone{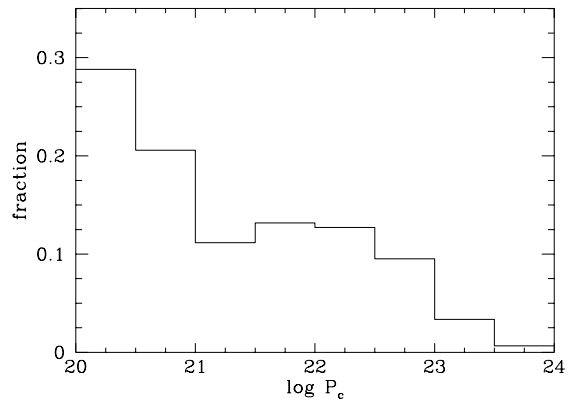}
\caption{The histogram of the values of $\log P_c$ (where
$P_c$ is the radio core power in W/Hz) for all 149 Seyfert galaxies.
The $\log P_c$--distribution function is calculated by means of the
Kaplan--Meier estimator.}
\end{figure}

The radio cores of Seyfert galaxies are orders of magnitude in power
stronger than the radio cores of normal "quiescent" spirals (e.g.,
$\log P_c<16$ W/Hz at similar frequencies for the Milky Way (Sgr
$A^{*}$) and for M31; see, e.g., the reviews by Lo, 1994, and by
Kormendy \& Richstone, 1995). They are also more powerful (by more
than one order of magnitude) than the (rarely detected) radio cores of
samples of nearby, non--Seyfert spirals, including objects with
starburst nuclei (e.g., Colina \& P\'erez--Olea, 1993; Sadler et al.,
1995). In particular, we note that the 25th percentile of the $\log
P_c$--distribution of Sadler et al.'s (1995) PTI core data for 29
spirals is less than 19.5 (W/Hz) at 5GHz (i.e., $\sim$19.7 (W/Hz) at
2.3 GHz, for a spectral index n=-0.7).

The power of Seyfert radio cores is rougly comparable to that of radio
cores detected in luminous IRAS galaxies, with starburst or AGN
nuclear spectra (Lonsdale et al., 1993), and in many low--redshift
($z<0.1$), optically--selected radio--quiet objects of the Bright
Quasar Sample (see, e.g., Kellermann et al., 1989; Lonsdale, Smith \&
Lonsdale, 1995).  These two surveys, having a lower sensitivity
threshold, contain a lower fraction of censored fluxes than Roy et
al.'s (1994) sample.

On the other hand, a comparison with Slee et al.'s (1994) wide sample
of early--type galaxies of low total radio power (denoted as subsample
A by the authors) reveals that these objects typically host stronger
radio cores than our Seyferts (with a Kaplan--Meier estimate of the
median of $\log P_c$=21.3 (W/Hz) at 2.3 GHz and a large percentage
($\sim$65\%) of core detections). If the comparison is limited to
subsets of nearby objects (distance D$<$100 Mpc), this difference
increases. Thus, our analysis strengthens Sadler et al.'s (1995)
suggestion (based on a much smaller data set) that radio cores are
more prominent in early--type than in Seyfert objects. Furthermore,
their multifrequency PTI observations revealed that the radio cores
detected in spiral (mostly Seyfert) galaxies usually have steep radio
spectra (with a median spectral index n=-1.0 between 2.3 GHz and 8.4
GHz), in contrast with the flat radio spectra of Slee et al.'s (1994)
early--type galaxies (median n=+0.3).

\subsection{Radio cores and Seyfert types.} 

In general, we rely on the Seyfert classification given by the
authors, with the exception of a few objects (NGC 2992, Tol 113, UGC
12138) which we class as S2 galaxies (NGC 2992 was classed as type 1.9
by Durret (1989) and Whittle (1992a); Tol 113 was given a subtype 1.9
by Whittle (1992a); UCG 12138 was found to be a subtype 1.8 by
Osterbrock \& Mantel (1993)). Moreover, we consider separately Mrk 516
and Mrk 883, which are probably Seyfert/LINER transition cases (see
Rudy \& Rodriguez--Espinosa, 1985).

Roy et al. (1994) employed the "difference-of-two-proportions" test
with the Yates continuity correction to quantify the difference in
detection rates between far-infrared (FIR)--selected S1s and
S2s. Within their far-infrared (FIR)--selected sample, where the two
Seyfert types have the same distance distribution, the authors found
the difference to be significant to the 98.8\% confidence level. The
combined optical and mid-infrared (MIR)-- selected sample may also
show a lack of detected radio cores in S1s, but that effect may be
simply explained by the fact that S1s are typically more distant than
S2s.

We wish to apply the methods of survival analysis in order to evaluate
the difference between the radio core emission of S1s and S2s. In this
case, we wish to compare the distribution functions of a given
quantity (the radio core power) for two different samples of objects
(S1 and S2) by testing the "null" hypothesis that the two independent
random samples are randomly drawn from the same parent
population. This is accomplished by using two versions of Gehan's test
(one with permutation variance and the other with hypergeometric
variance, hereafter called $G_1$ and $G_2$, respectively), the logrank
($L$), the Peto-Peto ($P_1$) and Peto-Prentice ($P_2$) tests. These
two-sample tests give the probability $p$ that two samples as different 
as these would be drawn from one parent population.
The statistical significance of the
difference between the distributions of two datasets is at the $100
(1-p)$\% level. The two-sample tests differ in how the censored points
are weighted and consequently have different sensitivities and
efficiencies with different distributions and censoring patterns.
Since two--sample tests seek only to compare two samples rather than
to determine the true underlying distribution of a given variable,
they do not require that the censoring patterns of the two samples be
random.

If we applied the five two--sample tests to the distributions of the
radio core powers $P_c$ of S1 and S2 objects, we would obtain only an
upper limit of $p$ (an average of $p=0.06$), because S1s are typically
more distant than S2s and, hence, tend to be brighter (or to have
greater upper limits on luminosities).  In order to remove this
distance effect in the comparison of subsets of objects lying at
different typical distances, we consider the logarithmic relation
\begin{equation}      
\log P_c = 17.078 + \log F_{min} + 2 \log D,
\end{equation} 

which gives the minimum value of $\log P_c$ (with $P_c$ in W/Hz)
corresponding to the $5\sigma$ sensitivity threshold of $F_{min}=3$
mJy for a detected object at luminosity distance $D$ (in
Mpc). Obviously, detections and upper limits on $P_c$ cluster near
this relation in the $\log P_c-\log D$ diagram; this diagram is
illustrated in Figure 2.
We use this relation, which illustrates the sensitivity threshold of
the radio survey, in order to evaluate the {\it observed} minus {\it
calculated} value of $\log P_c$, i.e. $\Delta(\log P_c)$, for each
object. In this manner we obtain $\Delta(\log P_c)$--distributions
which are free from distance selection effects (see Figure 3).
Applying the two--sample tests to the $\Delta(\log
P_c)$--distributions of S1s and S2s (with N= 57 S1 and $N_{ul}$= 43
upper limits; N=90 S2 and $N_{ul}$=48), we find that they differ at
the average level of p=0.006 (p=0.006, p=0.006, p=0.010, p=0.007,
p=0.003 according to the $G_1$, $G_2$, $L$, $P_1$, $P_2$ tests,
respectively). Thus, our analysis improves the statistical
significance of the difference. The $\Delta(\log P_c)$--distributions
are characterized by the medians of -0.33 and -0.15 for S1s and S2s;
this suggests a typical difference of a factor of $\sim1.5$ between
the radio core powers of the two types of Seyferts.

\begin{figure}[h]
\plotone{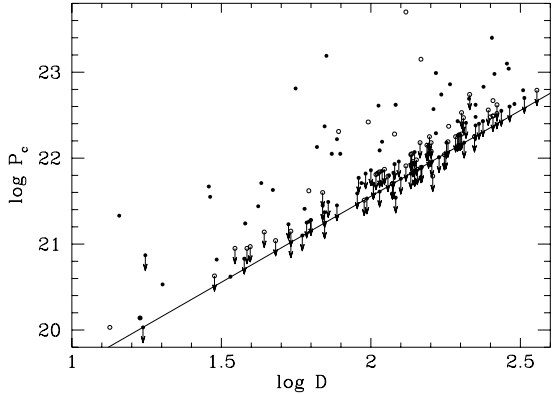}
\caption{The $\log P_c$--$\log D$ plot for 149 Seyfert
galaxies; $P_c$ is the radio core power (in W/Hz), D is the distance
(in Mpc).  Upper limits on $P_c$ are indicated by arrows. Different
symbols denote S1s (open circles), S2s (dots), and objects of
uncertain type (crosses). The straight line given by eq.  (1) in the
text is shown.}
\end{figure}

Within the standard unification schemes of Seyfert galaxies, if the
radio cores were located near obscuring NLR clouds, we would expect
a difference of a factor two between the radio core powers of S1s
and S2s, which is not too far from the difference we find.
As a matter of fact, if the NLR clouds lie closer to the nucleus
than do radio cores, when observing S1s along the axis of the NLR,
radio emission from radio cores on the far side of the nucleus would
be blocked from view by the NLR clouds with which they are associated,
and so only the components on the near side would be visible in the
radio. On the other side, S2s are viewed perpendicular to the axis
of the NLR and so our view of the radio cores would not be obscured
by the NLR clouds (see Figure 5b in Roy et al., 1994). Then the
radio power of S1 cores would appear to be typically half that
of S2 cores (as already suggested by Roy et al., 1994).
A similar mechanism would operate if the radio cores were on the
nuclear side of the NLR clouds or in a mixture of the two cases.

On the other hand, if the radio cores were located within (or
near) the BLR, a large covering factor of the NLR (closer to 0.5 than
to 0.01) by optically thick clouds is required to give rise to some
detectable difference in $P_c$ between S1s and S2s. Such large
covering factors disagree with the expectations of standard
photoionization models of the NLR, but are consistent with recent
views which take into account the presence of dust in the NLR gas;
dust tends to suppress narrow line emission, which implies that
photoionization models largely underestimate the covering factor (Netz
\& Laor, 1993).

\begin{figure}[h]
\plotone{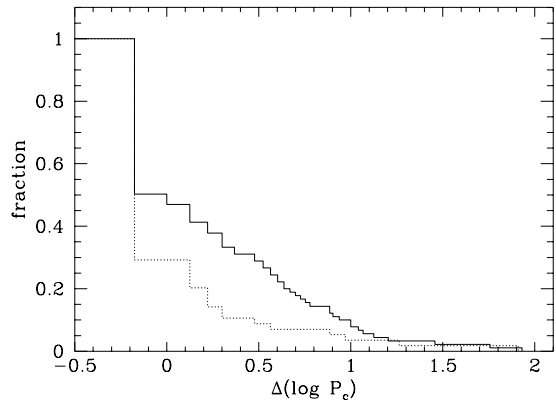}
\caption{The cumulative distribution function of
$\Delta(\log P_c)$ for Seyfert types 1 (dotted line) and types 2
(solid line). $\Delta(\log P_c)$ is the difference between the
observed value of the logarithmic radio power and the calculated value
(corresponding to the sensitivity threshold of the radio
survey).}
\end{figure}

\subsection{Radio cores and host galaxy morphologies.}

We have taken the morphologies of Seyfert galaxies generally from
Whittle (1992), the RC3 catalogue (de Vaucouleurs et al., 1991), the
catalogue by Lauberts \& Valentijn (1989), MacKenty (1990), Kirhakos
\& Steiner (1990), Vader et al. (1993), and the catalogue by
Lipovetski, Neizvestny \& Neizvestnaya (1987). We have simply
classified our objects as E/S0 (N=31) (whenever no trace of spiral
structure is observed), S (spirals) (N=72 objects, which are generally
of subtypes earlier than Sbc), and S/S0 (N=4) (in a few cases of
uncertain classification).
 
Many Seyfert galaxies, visually classified as ellipticals, are likely
to be really lenticulars, since recent attempts (Zitelli et al., 1993;
Granato et al., 1993; Kotilainen \& Ward, 1994) to decompose Seyfert
images into three major light components (bulge, disk, and nucleus)
have revealed that a disk is required in almost all cases. Many
objects lack morphological classification; this is also due to the the
widespread presence of amorphous, disturbed and peculiar morphologies
in Seyfert galaxies (e.g., Adams, 1977; MacKenty, 1990).  Amorphous,
very peculiar, and compact morphologies are left unclassified.
Although some amorphous objects may have been inconsistently
classified in the literature as E/S0, they tend to have less dominant
disks than spirals (McKenty, 1990).

Applying the above-mentioned two-sample tests to the $\Delta(\log
P_c)$--distributions relative to subsets of galaxies of different
morphological types, we find that compact radio cores tend to be more
frequently detected in E/S0 objects than in objects having other
morphologies, typically at a significance level close to $2\sigma$.
Table 1 gives the numerical outcomes, namely the total number of
objects N, the number $N_{ul}$ of upper limits on $P_c$, and the
probabilities $p$ that the two subsamples come from the same
underlying distribution according to the application of the five
two--sample tests on some subsample pairs. The effect, which indicates
that E/S0 galaxies have a radio core emission that is more powerful
than other objects typically by a factor of two, becomes marginally
significant in subsamples of relatively nearby objects (e.g., with
distance $D<150$ Mpc), because of poorer statistics. On the other
hand, in this respect no difference is observed between spirals of
earlier (S0/a, Sa, Sab) and later (Sb, Sbc, Sc) subtypes.

Our finding, which is not due to different proportions of
morphological types in S1 and S2 objects, reflects the presence of
some radio--loud objects (e.g., 3C120, 3C327) in our sample;
radio--loud objects are known to be always associated with early--type
galaxies (e.g., the review by Wilson, 1992). Our result appears to be
in line with the fact that in radio surveys of generic samples of
bright normal galaxies, compact radio cores are more common in E and
S0 galaxies than in later types (see, e. g., the review by Roberts \&
Haynes, 1994). Moreover, it is also consistent with the fact that
generic samples of early--type galaxies show more prominent compact
radio cores than Seyfert objects (see end of \S 2.1).

\subsection{Radio cores and optical emission.}

Searching for correlations between the radio core emission and the
galaxy optical luminosity, we analyzed the correlation between the
radio core logarithmic fluxes $\log F_c$ and the corrected total blue
magnitudes $B^{0}_T$, as well as the correlation between the
corresponding values of $\log P_c$ and absolute magnitude $M_B$.

In general, we take $B^{0}_T$ directly from the RC3 catalogue. If it
is not available there, we take the uncorrected $B_T$ magnitudes from
Lauberts \& Valentijn (1989), Whittle (1992a), Heisler \& Vader
(1994); for Mk 291 and Mk 841 we take the B magnitudes derived by
MacKenty (1990) for 30 kpc metric apertures (which are much greater
than the galaxy sizes corresponding to the isophotal diameters at the
25 mag/arcsec$^{2}$). We evaluate the total magnitudes of 3C120 and
Fairall 49 by simply summing up the non-stellar nuclear component and
the galaxy component as separately estimated by Kotilainen \& Ward
(1994). We correct these $B_T$ magnitudes for Galactic absorption,
internal aborption, and K dimming using the precepts given in RC3. Of
the three corrections, the K-dimming correction is least important,
whereas the corrections for internal absorption and Galactic
absorption are typically about equally important in spiral objects. In
cases of unknown morphological type we assume a Sa spiral morphology
(T=1 according to the the RC3 code), since most Seyfert galaxies are
early-type spirals. We also take the $B_T$ magnitudes (corrected for
Galactic absorption and K--dimming) as derived by Heisler \& Vader
(1994) for a few objects in common with our list.

In general, we analyze the significance of the correlations between
two "uncensored" variables by computing the two non-parametric rank
coefficients, Spearman's $r_s$ and Kendall's $r_k$ (see, e.g., Kendall
\& Stuart, 1977). In our case at least one variable ($F_c$ or $P_c$)
is censored. In testing the significance of correlations between the
variables $x$ and $y$ containing censored data, we evaluate the
probability $p$ that the two variables are independent, relying on the
Cox proportional hazard model (which allows censored data for only one
variable), on the generalized Kendall rank correlation statistics, and
on the generalized Spearman rank order correlation coefficient. The
last two methods also allow simultaneous censoring in both variables,
but Spearman's is known to be inaccurate for small samples ($N<30$).
These three correlation tests are hereafter referred to as $C$, $K$,
$S$, respectively. These correlation tests do not strictly require
that the censored points be randomly distributed, but perform better
if the censored points are not localized in a particular quadrant of
the $xy$ plane. In several cases, for simplicity's sake we shall
mention only the mean probability $p$ which results from the
application of the three correlation tests.

In order to examine the relationship between the radio core emission
and emissions in other wavebands, we shall deal with correlations both
between fluxes and between luminosities. Flux--flux plots will reveal
the intrinsic relationship between luminosities under limited
circumstances (in the case of uncensored data and linear
relationship). Therefore, we will study also luminosity--luminosity
correlations, but we will keep in mind that the use of luminosities
instead of fluxes tends to introduce a distance bias to the data, as
luminosities are correlated with distances in flux--limited
samples. In the case of significant total luminosity--luminosity
correlations, we shall attempt to estimate the influence of the
distance effect on the correlations between luminosities in order to
draw reliable conclusions on the existence of a physical relationship
between emissions in two wavebands.

Basically, we shall do this by applying the new test for partial
correlation in the presence of censored data (hereafter $AS$ test)
developed by Akritas \& Siebert (1995), who have extended to censored
data the concept of Kendall's partial rank correlation with uncensored
data (e.g., Siegel, 1956). The resulting partial correlation
coefficient gives a measure of the correlation between two data sets
(e.g., luminosities in two wavebands, in our case) independently of
their correlation with a third data set (the galaxy distances, in our
case). Thus, the $AS$ partial correlation test is a good approach to
free the luminosity--luminosity correlation of the individual
luminosity--distance correlations. We use the AS test also for partial
correlation analysis of uncensored data sets.

Applying the $C$, $K$, $S$ correlation tests, we find no significant
$\log F_c$--$B^{0}_T$ correlations for S1(N=39, $N_{ul}$=27), S2
(N=51, $N_{ul}$=25), and all objects (N=92, $N_{ul}$=54). The study of
the $\log P_c$--$M_B$ total relation confirms the absence of a total
correlation for S2s, but yields a fairly strong total correlation for
S1s (with a mean $p=0.02$) together with a marginal correlation for
the whole sample (with a mean $p=0.08$). In order to explore whether
these two correlations can be totally induced by a distance bias we
apply the $AS$ partial correlation test to the $\log P_c-M_B$
relation. We find no significant partial correlation between $\log
P_c$ and $M_B$, which means that the total correlation between the two
quantities is simply an artifact of the distance bias.

In order to verify this conclusion, we have taken the homogeneous
(albeit not very accurate) set of total V magnitudes (corrected for
foreground extinction and galaxy inclination) as estimated by De Grijp
et al. (1992) for the objects common to their sample. Again, we find
no significant $\log F_c$--$V$ correlation for S1 (N=30, $N_{ul}$=23),
S2 (N=65, $N_{ul}$=34), and all objects (N=95, $N_{ul}$=47), but a
correlation between intrinsic luminosities ($\log P_c$ and $M_V$) for
S2s (with a mean $p=0.03$) and for all objects (mean $p=0.04$).  By
applying the $AS$ test we verify again that these total correlations
are simply spurious distance effects.

For the S2 galaxies, which generally have a small non-stellar
(nuclear) light component, we can further try to estimate the bulge
magnitude simply by subtracting the contribution of disk emission. We
have given a rough estimate of the disk contribution by relying on the
empirical mean relation between the bulge-to-disk light ratio and the
morphological type as determined by Simien \& de Vaucouleurs (1986)
from bulge/disk decomposition of nearly one hundred galaxies. Their
sample contains several Seyfert galaxies and these do not deviate
systematically from the mean relation. Explicit bulge/disk light
decompositions have not been used, since the scatter in the
bulge-to-disk light ratio versus morphological type is mainly due to
errors in photometric decomposition (see, e. g., Giuricin et al.,
1995a). According to the mean relation derived by Simien \& de
Vaucouleurs (1986), we adopt the following conversion $\Delta m$ (in
magnitude) from bulge+disk magnitudes to bulge magnitudes as a
function of the Hubble morphological stage T (coded as in RC3): 0 mag
for ellipticals, 0.47 mag for T=-3, 0.61 for T=-2, 0.73 for T=-1, 0.86
for T=0, 1.02 for T=1 (Sa) , 1.23 for T=2, 1.54 for T=3 (Sb) , 1.97
for T=4, 2.54 for T=5 (Sc); we adopt $\Delta m$=0.6 for generic S0,
$\Delta m$=0.5 for E/S0 objects, $\Delta m$=1.0 for generic spirals or
for objects with unknown morphology. For S2 objects with available
$B_T$ (N=51, $N_{ul}$=25) or available V (N=65, $N_{ul}$=34), we find
that $F_c$ is unrelated to the apparent magnitude of the bulge. Also
$P_c$ is unrelated to the absolute magnitude of the bulge, when the
distance effect is removed.

We also wish to explore the relation between the radio core emission
and the non-stellar optical B, V, R magnitudes evaluated by some
authors through a subtraction of the host galaxy stellar component,
whose contribution is important even in small nuclear apertures for S1
objects.  In the choice of data we give preference to the CCD imaging
studies by Granato et al. (1993) for an optically-selected (mostly
UV-excess selected) sample of S1 and those by Kotilainen, Ward \&
Williger (1993) for a hard X-ray selected sample (see also Kotilainen
\& Ward, 1994). In case of several entries for one object we take the
average of the fluxes.  Granato et al. (1993) listed magnitudes
corrected for Galactic absorption.  We correct Kotilainen et al.'s
(1993) nuclear magnitudes for Galactic absorption as in RC3 , for a
standard extinction law ($A_V= 0.75 A_B$; $A_R=0.74 A_V$). There are
N=25 objects (with $N_{ul}$=15 upper limits on $F_c$) with known
non-stellar B magnitudes in our sample (N=26 with $N_{ul}$=16 for the
V and R data). They are mostly S1 objects.

For S1 and all objects, we detect neither significant correlations
between $F_c$ and the apparent nuclear magnitudes (in all three bands)
nor significant partial correlations between $P_c$ and the
corresponding absolute nuclear magnitudes. This result does not change
if we add a few nuclear magnitudes (corrected for Galactic absorption)
that we derived from crude estimates of magnitudes or fluxes available
in the literature (B=15.6 for Mk 1148 (Smith et al., 1986), R=15.6 for
MCG-6-30-15 (McAlary \& Rieke,1988), B=14.4 for II Zw 136 (Smith et
al., 1986)).

For N=25 objects ($N_{ul}$=14) with known $B^{0}_T$ and non-stellar B
magnitudes, we can first subtract the non-stellar contributions from
the total emission in blue light and then the disk contribution
obtained by means of the above--mentioned $\Delta$m--values. These
objects are mostly S1s (N=20, $N_{ul}$=12). Again, $F_c$ and $P_c$
turn out to be unrelated to their total and disk stellar
emissions. The addition of a few S2s with available non--stellar B
magnitudes leads to a significant $P_c$--absolute magnitude
correlation, which can be simply explained by a selection effect (for
a given total luminosity, S2s, hosting much fainter optical nuclei,
have brighter bulge and disk luminosities than S1s).

We conclude that the radio core emission is unrelated to the total
optical luminosity of the host galaxy, to the total and bulge stellar
luminosity and to the non--stellar nuclear luminosity. We have
verified that this holds true no matter whether we remove E/S0
galaxies from all samples or whether we restrict ourselves to
uncensored radio data only.

Thus, in Seyfert galaxies the behaviour of radio core emission clearly
differs from that of total radio emission and that of central radio
emission (coming from areas of a few arcsecs in size); both emissions
appear to be related to the total optical luminosity (e.g., Edelson,
1987) and especially to the bulge optical luminosity (e.g., Whittle,
1992c).

It is interesting to note that also the compact ($<$0.03 arcsec) radio
cores detected at 2.3 and 8.4 GHz with the Parkes--Tidbinbilla
Interferometer in generic samples of early--type galaxies (Slee et
al., 1994) are unrelated to the $M_B$ absolute magnitudes. We have
checked this by applying the $AS$ correlation test to the radio core
powers (interpolated at 5 GHz) and $M_B$--values, as tabulated by Slee
et al. (1994), for their subsets of galaxies of low and high radio
power (N=62, $N_{ul}$=19 and N=85, $N_{ul}$=25, respectively).

Our result appears to be particularly robust for the total optical
luminosity (because of the fairly large data sample) and less reliable
for the non--stellar nuclear light, because of small--number
statistics. If we want to regard Seyfert nuclei simply as scaled--down
versions of quasars, we would expect a radio--optical correlation such
as that noted by Lonsdale et al.  (1995) for the fairly compact radio
structures detected by Kellermann et al.  (1989) in their 5 GHz VLA
survey of the Bright Quasar Sample (at $\sim$0.5" and $\sim$18"
resolutions) . However, by applying the $AS$ correlation test to the
luminosities tabulated by Lonsdale et al. (1995), we find that the
optical--radio core correlation is not statistically significant for
the subsample (N=23, $N_{ul}$=7) of low--redshift ($z\leq$0.10)
objects. One has to almost double the number of objects (increasing
the limiting $z$--value up to $z\sim$0.2) in order to reach a
significant correlation at a $\geq$2$\sigma$ level. But the compact
radio sources detected in distant quasars can hardly be compared to
our Seyfert radio cores, which have physical sizes smaller by one
order of magnitude (or more).

\subsection{Radio cores and near-infrared emission.}

We have explored the relations between the radio core emission and the
non-stellar (nuclear) near-infrared (NIR) J, H, K magnitudes. We took
these magnitudes directly from the results of the NIR (K--band)
imaging studies (complemented by NIR photometry) of Zitelli et
al. (1993) and Danese et al.  (1992), from the NIR imaging
observations carried out by Kotilainen et al.  (1992a, b), Kotilainen
\& Prieto (1995), and from the K imaging of the
spectroscopically--selected CfA Seyferts by McLeod \& Rieke (1995), in
order of preference. These observations allowed the authors to
separate the nuclear light from the stellar emission of the host
galaxy. For N=28, 27, 34 objects (with $N_{ul}$=17, 16, 21 ) -mostly
S1 - we notice marginal correlations between $F_c$ and the apparent
NIR magnitudes (for all objects, not for S1s alone) together with no
partial correlations between $P_c$ and absolute NIR magnitudes.

Spinoglio et al. (1995) have estimated total near-infrared magnitudes
for 27 Seyfert objects in common with our list. For S2s the authors
used RC3 growth curves to extrapolate from measures obtained in the
largest apertures to the total flux (as done for non-Seyfert
galaxies), but only when the largest apertures are so large that the
nuclear component no longer affects the shape of the growth curve. For
S1s, whose NIR light is much more strongly concentrated than any of
the normal galaxy growth curves, they used the galaxy/nucleus
decompositions mentioned in the last paragraph and summed up these two
components to obtain the total NIR flux.  We used the latter approach
to obtain the total flux in the K band for another 19 Seyfert galaxies
(not included in Spinoglio et al., 1995) with available galaxy/nucleus
decompositions. For a total of 23 S1s ($N_{ul}$=13), 23 S2s
($N_{ul}$=10), and S1+S2, we do not find any correlation between core
radio flux (luminosity) and total K-band flux (luminosity).

In conclusion, the radio core emission is unrelated to the total and
non--stellar nuclear NIR emissions. This holds true whether we drop
out E/S0 galaxies and or retain detected objects only.  As in the case
of the optical--radio correlation discussed in the previous
subsection, our result is more reliable for the total than for the
nuclear NIR luminosity, for which statistics are poor.
 
\subsection{Radio cores and mid--, far--infrared emissions.}

Mid--infrared (MIR) emission from Seyfert nuclei is substantially due
to non--stellar radiation. Nuclear radiation predominates in the
small-beam MIR measures and also in the large-beam IRAS MIR data,
whereas the bulk of the IRAS far-infrared (FIR) radiation is
contributed by the emission of galaxy disks (e. g.,
Rodriguez-Espinosa, Rudy \& Jones, 1987; Giuricin et al.,
1995b). Consulting basically the compilation by Giuricin et al.
(1995b) and the new observations by Maiolino et al. (1995), we have
gathered together ground-based small-aperture ($\lesssim10"$) measures
in the standard N band ($\lambda\sim10\mu$m).  We have also collected
IRAS MIR (at $\lambda\sim12\mu$m and $\sim25\mu$m) and FIR (at
$\lambda\sim60\mu$m and $\lambda\sim100\mu$m) measures, giving
preference to reference sources which report co-added survey data
(e. g., Rush, Malkan \& Spinoglio, 1993; de Grijp et al., 1992;
Mazzarella, Bothun \& Boroson, 1991; Sanders et al., 1989; Hill,
Becklin \& Wynn-Williams, 1988; Helou et al., 1988) and sensitive
pointed observations (Edelson \& Malkan, 1986) over catalogues which
report IRAS point source fluxes (e. g., Fullmer \& Lonsdale,
1989). The consultation of the "Catalog of Infrared Observations" by
Gezari et al. (1993) has been of valuable aid in compiling infrared
data from the literature.

We find that $F_c$ is positively correlated with the fluxes in the N
band and in four IRAS bands (hereafter $F_N$, $F_{12}$, $F_{25}$,
$F_{60}$, $F_{100}$) for all objects and for S2s because of the the
censoring pattern of the $F_c$-values of S2s (censored values of $F_c$
cluster in the region of low infrared fluxes). However, the
application of the $AS$ test to the luminosity-luminosity diagrams,
indicates that $P_c$ correlates fairly well with the (monochromatic)
luminosities $L_{12}$ (at the $\sim$ 98\% significance level) and
marginally with $L_{25}$, for all objects. If S1s and S2s are analyzed
separately, we still find marginal $\log P_c-\log L_{12}$ correlations
for S1s and S2s together with a moderate $\log P_c-\log L_{25}$
correlation (at the $\sim$95\% level) for S2s. Moreover, whenever we
restrict ourselves to objects detected both at radio and infrared
wavelengths, we still detect partial luminosity--luminosity
correlations for $L_{12}$ (N=41, $\sim$95\% level) and for $L_{25}$
(N=55, 99.7\% level). There are no partial luminosity--luminosity
correlations in the other wavebands (not even in the case of
uncensored data only).

Furthermore, we explore possible correlations of radio core emission
with the IRAS colours $F_{12}/F_{25}$, $F_{12}/F_{60}$,
$F_{25}/F_{60}$, $F_{60}/F_{100}$, omitting some galaxies which are
undetected in two adjacent IRAS bands. $F_c$ does not correlate with
the IRAS colours, whereas $P_c$ shows significant positive total
correlations with the ratios $F_{12}/F_{60}$, $F_{25}/F_{60}$, and
$F_{60}/F_{100}$. However, by
applying the $AS$ test (with the distance taken as the third variable)
we have verified that the latter correlation is totally induced by the
fact that distant and MIR--bright objects tend to have greater
$F_{60}/F_{100}$-values (see, e.g., Giuricin et al., 1995b). On the
other hand, we found that the relations $\log P_c$--$\log
(F_{12}/F_{60})$ and $\log P_c$--$\log (F_{25}/F_{60})$ are not
entirely induced by the positive relations between MIR luminosity and
$F_{MIR}/F_{FIR}$ colours, which are ascribed to a decreasing relative
contribution of the cool emission from galaxy disks in
high--luminosity AGNs (Giuricin et al., 1995b); even when the distance
effect is removed, they remain significant (to the 98\% and 97\%
confidence level, respectively, for all objects). In particular the
partial $\log P_c$--$\log (F_{12}/\log F_{60})$ correlation is
significant even for uncensored data only (N=54, $>$99.9\% level). The
two correlations are consistent with the fact that $P_c$ is related to
$L_{12}$ and $L_{25}$, but not to $L_{60}$.

Table 2 presents some results relative to the correlations between
radio core power and infrared luminosity or colour, i.e. the total
number of objects (N), the number of upper limits on $F_c$ ($N_{ul}$),
the number of upper limits on the infrared quantity ($N'_{ul}$), the
partial Kendall's $\tau$ correlation coefficient, the square root of
the calculated variance, $\sigma$, and the associated probability $p$
of erroneously rejecting the null hypothesis (i.e., no correlation)
(see Akritas \& Siebert, 1995). Figure 4 shows the $\log P_c-\log
L_{12}$ plot.
 
We conclude that radio core emission tends to increase with MIR
emission (which is essentially a property of the nuclear source) and
with the flattening of the FIR to MIR energy distribution, whilst it
is unrelated to the FIR emission (which comes mostly from the host
galaxy). The absence of correlations between $P_c$ and N-band infrared
luminosity might be due to small--number statistics.

\begin{figure}[h]
\plotone{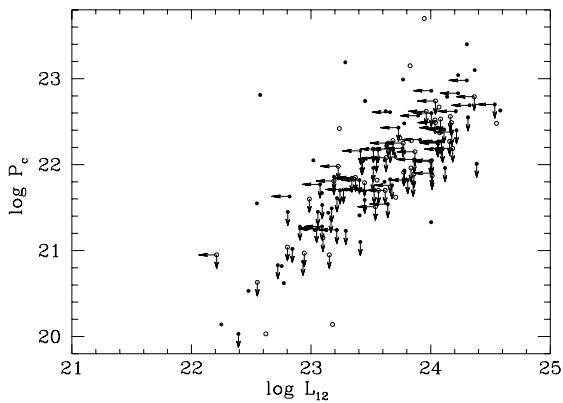}
\caption{The $\log P_c$--$\log L_{12}$ plot, where $P_c$
is the radio core power (in W/Hz), $L_{12}$ is the IRAS 12$\mu$m
luminosity (expressed in W/Hz). Symbols as in Figure 2.}
\end{figure}

Incidentally, Seyfert galaxies show a relation between the central
N--band emission and the (total and extended central) radio emission
(e.g., Ulvestad, 1986; Telesco, 1988); besides they appear to lie on
the same tight FIR--total radio correlation as normal spirals,
ultraluminous infrared galaxies, and radio--quiet quasars (e.g., Sopp
\& Alexander, 1991).

\subsection{Radio cores and X-ray emission.}

We wish to investigate the relation between the radio core emission
and the soft and hard X-ray emissions coming from various X-ray
surveys.  In the case of multiple entries for a galaxy, we adopt the
mean values of the X-ray fluxes $F_x$ published in a reference source.
We choose unabsorbed fluxes if available, although in some cases
(e.g. NGC 1068) standard corrections for flux absorption are
inadequate.

The first sample of soft X-ray fluxes (in the 0.5--4.5 keV band) is
based on the large compilation by Green, Anderson \& Ward
(1992). These authors compiled the IR and X-ray luminosities of a
large sample of normal galaxies, starburst galaxies, and AGNs,
observed with both the {\it Einstein Observatory} and IRAS
satellites. Almost all the X-ray measures were taken with the Imaging
Proportional Counter (IPC) aboard the {\it Einstein Observatory}
satellite, although a few were obtained with the High Resolution
Imager (HRI). The authors converted published X-ray luminosities to
the chosen bandpass (0.5--4.5 keV), when necessary, and they used
published fluxes obtained from X-ray spectral fits, when
available. From the tabulated luminosities (corrected for Galactic
absorption) and redshifts, we estimated the X-ray fluxes $F_x$ (or
upper limits) for 37 Seyferts (mostly S1) in common with our sample
(with $N_{ul}$=20 upper limits on $F_c$ and $N'_{ul}=1$ upper limit on
$F_x$).

Furthermore, we consider a second sample of soft X-ray data ({\it
Einstein Observatory} IPC fluxes) which include the unabsorbed X-ray
fluxes (in the 0.2--4 keV band) derived by Kruper, Urry \& Canizares
(1990) for 27 Seyferts (mostly S1) in common (with $N_{ul}$=15)
together with the X-ray fluxes (in the same band) or upper limits
derived by Fabbiano, Kim \& Trinchieri (1992) for another 5 Seyferts
(with $N_{ul}$=2,$N'_{ul}$=2).

The third sample of soft X-ray data comprises the {\it Einstein
Observatory} IPC fluxes (in the 0.16--3.5 keV) derived by Wilkes et
al. (1994) for 20 S1 objects in common (with $N_{ul}$=14).

Two sets of 2 keV fluxes were derived by Walter \& Fink (1993) from
the spectral analysis of the soft (0.1--2.4 keV) spectra of Seyferts
detected with the PSPC detector aboard ROSAT. They described the
spectra in two manners, namely using a hard X-ray power-law component,
a low energy absorption, both with and without the addition of a soft
X-ray component represented by a thermal bremsstrahlung model. In both
cases the 2 keV fluxes are mainly contributed by the hard X-ray
power-law component. There are N=15 ($N_{ul}$=10) objects in common.

The ROSAT All Sky Survey of IRAS galaxies (Boller et al., 1992)
provides a fifth sample of 19 objects (with $N_{ul}$=13) detected with
the PSPC detector.  We considered the tabulated fluxes in the 0.1--2.4
keV energy band.

Lastly, we construct a combined sample of soft X--ray data, taking the
1 keV fluxes (corrected for absorption) as reported in the catalog of
X--ray spectra of AGNs by Malaguti, Bassani \& Caroli (1994) for 44
objects ($N_{ul}$=30) detected with the IPC or SSS detectors aboard
{\it Einstein Observatory} and with the PSPC instrument aboard
ROSAT. In the case of multiple entries for a galaxy we adopt the mean
value of $F_x$.

For the six samples of soft X-ray data, we found generally no
flux--flux ($F_c-F_x$) correlations (except for the last sample). For
S1 and all objects, we never notice partial luminosity--luminosity
($P_c$--$L_x$) correlations (although the 14 uncensored data of the
last sample correlate at the $\sim$98\% significance level).

Let us turn to higher--energies X-rays, which, as compared to soft
X-rays, suffer less attenuation by interstellar matter and are better
indicators of AGN emission (on energy and rapid variability grounds).
Short- and long-term hard X-ray variability is a common phenomenon in
AGNs, but the fluxes do not generally vary by more than a factor of
two (see, e.g., Grandi et al., 1992).

The HEAO1-A1 ($\sim$2--20 keV) sample of identified AGNs, the
so--called LMA sample studied by Grossan (1992) has N=18 objects (with
$N_{ul}$=11) in common.  We took the 5 keV fluxes as derived by
Grossan (1992).

The GINGA ($\sim$2--10 keV) sample of AGNs has N=16 objects (with
$N_{ul}$=6) in common. We took the 2--10 keV fluxes mostly from Nandra
\& Pounds (1994) and from Williams et al. (1992) for II Zw 136, Awaki
et al. (1991) for IRAS 15091-2107 and Fairall 49, Koyama et al. (1989)
for NGC 1068.

The EXOSAT X-ray spectra of detected AGNs analyzed by Turner \& Pounds
(1989) provide a set of hard X-ray (2--10 keV) fluxes for 17 objects
in common, to which we add the EXOSAT data of Mk 841 and II Zw 136
(Saxton et al., 1993), ESO12--G21 (Ghosh \& Soundararajaperumal,
1992), Mk 1148 (Singh, Rao \& Vahia, 1991), Mk 618 (Rao, Singh \&
Vahia, 1992), IRAS 12495--1308 and IRAS 15091--2107 (Ward et al.,
1988), Mk 705 (Comastri et al., 1992).

Of the above--mentioned X-ray data samples, which yield no
$F_c$--$F_x$ correlations, only the widest one, the EXOSAT sample
(N=25, $N_{ul}$=14), shows a weak partial, luminosity-- luminosity
($P_c-L_x$) correlation (at the 94.5\% confidence level).

Lastly, we construct a combined sample of hard X-ray broad-band fluxes
by adding to the EXOSAT sample the GINGA 2--10 keV fluxes of Fairall
49 (Awaki et al., 1991), the {\it Einstein Observatory} MPC 2--10 keV
fluxes (or upper limits) of NGC 1566, NGC 4235, Mk 541 derived by
Halpern (1982), the ASCA GIS+SIS estimated flux of NGC 5252 in the
2--10 keV band (Cappi et al., 1995).

This combined sample (N=30, $N_{ul}$=17, N'ul=2) confirms the presence
of a partial $P_c$--$L_x$ correlation, at the 97.9\% confidence level
(see Table 3) for all objects (of which N=13 objects with uncensored
data correlate at a similar level of statistical significance). Figure
5 shows the $\log P_c$--$\log L_x$ plot.

To sum up, there is no sure evidence that the radio core emission is
related to the soft X--ray emission. But we can say that it is likely
to be related to the hard X--ray emission, since this comes out in the
widest relevant data samples.

\begin{figure}[h]
\plotone{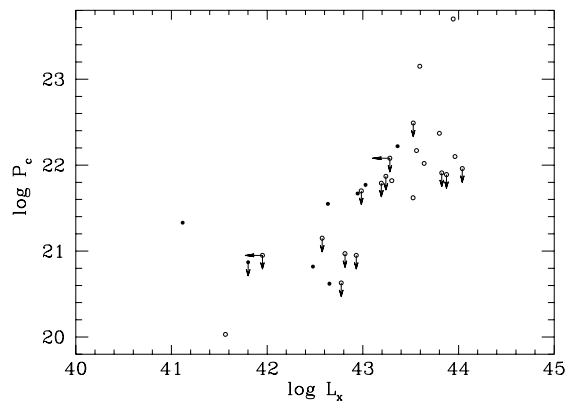}
\caption{The $\log P_c$--$\log L_x$ plot, where $P_c$ is
the radio core power (in W/Hz), $L_x$ is the X-ray luminosity
(expressed in erg/s) in the $\sim$2--10 keV energy band. Symbols as in
Figure 2.}
\end{figure}

Incidentally, several studies have established a correlation between
X-ray emission and radio emission in AGNs.  Radio-loud AGNs (quasars
and Seyferts) are generally believed to be stronger X-ray emitters
than their radio-quiet counterparts (e. g., Worral et al., 1987; Unger
et al., 1987; Kruper et al., 1990) and the proportion of radio-loud
AGNs appears to increase with X-ray luminosities (Della Ceca et al.,
1994).  Furthermore, the soft X-ray spectral slopes show some
dependence on radio power in quasars (e.g., Wilkes \& Elvis, 1987;
Baker, Hunstead \& Brinkmann, 1995).

\subsection{Radio cores and extended radio emission.}

In order to explore the relation between the radio core emission and
the total radio emission of Seyfert galaxies, we compile published
total radio fluxes giving preference to observations made at lower
resolution and at frequencies close to 2.3 GHz.  According to
Edelson's (1987) estimates, the total radio emission of Seyfert
galaxies includes at least a $\sim$20\% contribution from the disk of
the underlying galaxy.  We take the total fluxes $F_T$ of detected
Seyfert galaxies from Wright (1974), Wilson \& Meurs (1982), Gioia et
al.  (1983), Tovmassian et al. (1984), Ulvestad \& Wilson (1984a,b),
Unger et al. (1986, 1987), Harnett (1987), Burns et al. (1987),
Edelson (1987), Condon (1987), Antonucci \& Barvainis (1988), Ulvestad
\& Wilson (1989), Kellermann et al. (1989), Unger et al. (1989),
Condon et al. (1990), Condon \& Broderick (1991), Klein, Weiland \&
Brinks (1991), Gregory \& Condon (1991), van Driel, van den Broek \&
de Jong (1991), Neff \& Hutchings (1992), White \& Becker (1992),
Cram, North \& Savage (1992), Miller, Rawlings \& Saunders (1993),
Vader et al. (1993), Slee et al.  (1994), Gregorini et al. (1994),
Bicay et al. (1995). We converted the fluxes measured at different
frequencies (mostly at $\sim$1.4 GHz and $\sim$5 GHz) to 2.3 GHz with
the adoption of a spectral index $n=-0.7$.

We find total radio flux data for N=92 objects (with $N_{ul}$=55 upper
limits on $F_c$ and $N'_{ul}$=13 upper limits on $F_T$). As expected,
for objects with detected radio cores, the total radio flux $F_T$
converted to 2.3 GHz is in general, within the errors, not smaller
than the radio core flux $F_c$, with the notable exception of Mk 841,
for which Roy et al.  (1994) listed a radio core flux ($F_c$=55 mJy)
much greater than the total flux measured by Kellermann et al. (1989)
with the VLA in the D configuration at 5 GHz ($F_T$= 1.5
mJy). Moreover, in 15 cases, "censored" or "uncensored" values of
$F_T$ are smaller than the corresponding "censored" values of $F_c$,
mostly because of the limited sensitivity of Roy et al.'s (1994)
survey.

Wondering how much of the total radio flux comes from the compact
radio core itself, we calculate the Kaplan--Meier distribution of
$\log (P_c/P_T)$ for objects with detected $F_T$.  The distribution
function of $\log (P_c/P_T)$ has a median of -0.76, together with 75th
and 25th percentiles of -1.49 and -0.41. The medians are -0.90 and
-0.73 for S1s and S2s, respectively, but they are not statistically
different (because of small-number statistics). We have checked that
the median and the two percentiles do not change appreciably if we
replace some unreasonable, positive values of that quantity (see
previous paragraph) with zero.  Hence, we can say that the compact
cores contribute, on average, a relatively small fraction ($\sim$17\%)
of the total radio emission in Seyfert galaxies.

We have verified that greater fractions ($\sim$22\% and $\sim$56\%)
are contributed by the compact cores observed in Slee et al.'s (1994)
low-power sample of early--type galaxies (at similar resolution) and
by those observed in Lonsdale et al.'s (1995) QSO sample (at lower
resolution). The corresponding median contribution is smaller
($\sim$5\%) in Lonsdale et al.'s (1993) VLBI experiment, which has,
however, higher sensitivity than Roy et al.'s (1994) survey. The
median contribution is even smaller ($\sim$1\%) in Sadler et al.'s
(1995) sample of non--Seyfert spirals (observed at similar resolution
and sensitivity).

At first glance, there seems to be no significant partial correlation
between the radio luminosities $P_c$ and $P_T$ corresponding to the
nominal values of $F_c$ and $F_T$, in spite of the correlations
between fluxes and between uncensored values. But we have recognized
that this apparent negative result is essentially due to the presence
of several (15) cases in which a fairly high "censored" value of $P_c$
is accompanied by a lower, "censored" or "uncensored" value of
$P_T$. If we reasonably set the "censored" value of $P_c$ equal to the
corresponding value of $P_T$ in all these 15 cases, we find a
significant partial $P_c$--$P_T$ correlation, at the $\sim$~96\%
level, for S2 and all objects. Table 4 lists the relevant results
(symbols as in Table 2).
Subtracting the core flux from the total radio flux, we can perform
a similar analysis of  the correlation between $P_c$ and $(P_T-P_c)$.
In this case, when $P_c$ is censored, we simply take an upper limit
of $(P_T-P_c)$ equal to the corresponding value of $P_T$.
We find that $P_c$ is related also to the extended radio power $(P_T-P_c)$.

Our result is consistent with what is observed in many samples, which,
being less heavily censored, generally show a good correlation between
radio core and total emissions (e.g, Slee et al.'s (1994) early--type
galaxies, Lonsdale et al.'s (1995) QSOs, Neff \& Hutchings' (1992)
luminous IRAS galaxies), in agreement with results concerning powerful
radio galaxies (e. g., Fabbiano et al., 1984).  On the other side,
there is no such correlation within Lonsdale et al.'s (1993) luminous
IRAS galaxies and Sadler et al.'s (1995) heavily censored sample of
non--Seyfert spirals.

In order to calculate the linear regression line in the case of
censored data, we use Schmitt's method (e.g., Isobe et al., 1986),
which works even when censoring is present in both variables.  We
found $P_c \propto P_T^{1.0\pm 0.1}$ for all objects ($P_c\propto
P_T^{0.9\pm 0.1}$ for S1s and $P_c\propto P_T^{1.1\pm 0.1}$ for
S2s). If we drop out the "censored" values of $P_T$ and repeat the
linear regression analysis using also the parametric and
non--parametric EM algorithms, which work only when only one variable
is "censored", we confirm the above--mentioned result.  The value of
the slope we find is intermediate between the steep slope
(1.9$\pm$0.7) found by Sadler et al. (1995) for their spiral sample
and the typical value (0.7) for high- and low-radio power early--type
galaxies (e.g., Slee et al., 1994). However, if we repeat the linear
regression analysis for E/S0 Seyferts only (N=20 with $N_{ul}$=9,
$N'_{ul}=1$), we find a slope of 0.8$\pm$0.2, which is consistent with
the value of 0.7 mentioned above.
 
We have inspected the characteristics of the inner radio structure on
arcsec-- or subarsec--angular scale for 59 objects, for which relevant
information is available in the literature (i.e., in the references
already cited in this subsection and in Norris (1988), Wilson \&
Tsvetanov (1994), Kukula et al. (1995)). We have found that $P_c$ is
unrelated to the morphologies of the inner radio structures classified
as linear, diffuse, slighltly resolved, unresolved, according to the
classification introduced by Ulvestad \& Wilson (1984a).

Moreover, there are 35 objects for which high-resolution ($<$1")
MERLIN and mostly VLA published observations have detected an
unresolved nuclear component (with a radio size ranging from
$\lesssim$0.8" to $\lesssim$0.1"), which is often embedded in a
diffuse component or is a part of multiple and elongated
structures. Converting the powers of these nuclear radio components
(hereafter $P_n$), mostly measured at 5 GHz, to 2.3 GHz (with the
adopted n=-0.7), we evaluate the Kaplan--Meier $\log
(P_c/P_n)$--distribution (with N=35, $N'_{ul}$=16). We derive a mean
of -0.20$\pm$0.10, a median of -0.26, and a 75th (25th) percentile of
-0.60 (-0.05). From the fact that this distribution is shifted towards
negative values (at a $\sim2\sigma$ level), we argue that most objects
have extended radio structure on a subarsec--scale. Our results
strengthen those of Sadler et al.'s (1995) ones, which are based on a
smaller Seyfert sample and refer to nuclear components on larger
scales ($\sim$1--5 arcsec), observed with the VLA at 5 GHz. The same
authors have already recognized that the inner radio structures of
Seyfert galaxies do not look like those of Slee et al.'s (1994)
early--type galaxies, where compact radio cores dominate nuclear
components on an arcsec scale, observed at 5 GHz with the VLA or the
Australian Telescope (AT).

\subsection{Radio cores and emission line properties.}

Searching for correlations with radio core emission, we consider the
measures of [OIII]$\lambda 5007$\AA $\!$ narrow emission line flux
available in the literature, as compiled mostly by de Grijp et
al. (1992), Whittle (1992a), Dahari \& De Robertis (1988a) (in this
order of preference). Besides the fact that line fluxes correlate
fairly strongly with $F_c$, we detect also a weak partial correlation
(at the $\sim$93\% level) between the corresponding luminosities of
all objects (the correlation between N=50 detected objects reaches the
$\sim$96\% level).

\begin{figure}[h]
\plotone{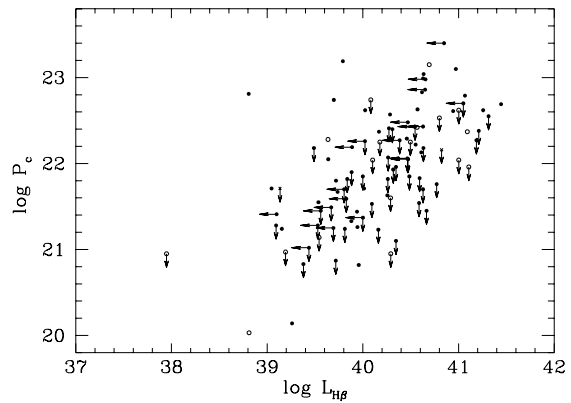}
\caption{The $\log P_c$--$\log L_{H\beta}$ plot, where
$P_c$ is the radio core power (in W/Hz), $L_{H\beta}$ is the
luminosity of the $H\beta$ narrow emission line (in erg/s). Symbols
as in Figure 2.}
\end{figure}

Furthermore, we consider the fluxes (or upper limits) of the narrow
and broad components of the $H\beta$ emission line as given mostly in
the three above--mentioned papers. We detect good partial correlations
(at the $\sim3\sigma$ level) between $P_c$ and the luminosity of the
$H\beta$ narrow line, $L_{H\beta}$, for S2 and all objects (although
fluxes and uncensored data are unrelated). Table 5 lists the relevant
outstanding results of our partial correlation analysis (symbols as in
Table 2).  Figure 7 shows the $\log P_c$--$\log L_{H\beta}$ plot for
all objects.

Moreover, we consider the [OIII]$\lambda5007$\AA $\!$ narrow emission
line widths measured at 50\% and 20\% levels with respect to the peak
intensity. We take the former quantity mostly from Nelson \& Whittle
(1995), Whittle (1992a) and in a few cases from Dahari \& De Robertis
(1988a) and Vader et al. (1993); for the latter quantity we rely on
the compilations by Nelson \& Whittle (1995) and Whittle (1992a). We
found that the two line width parameters are unrelated to $P_c$ (for
N=63, $N_{ul}$=36 and N=43, $N_{ul}$=24, respectively), even if we
drop out E/S0 galaxies or undetected objects.

For S2 objects, where the narrow components of Balmer lines are best
determined, we find no correlations between $P_c$ (or $F_c$) and an
indicator of optical excitation such as the narrow line ratio
[OIII]$\lambda5007$\AA/$H\beta$ (for N=84,$N_{ul}$=29 upper limits on
$F_c$, $N'_{ul}$=6 upper limits on $H\beta$ flux). Neither did we find
correlations between $P_c$ (or $F_c$) and the ratio
[NII]$\lambda6583$\AA/$H\alpha$ (for N=72, $N_{ul}$=39 upper limit on
$F_c$, $N'_{ul}$=1 upper limit on line ratio); almost all the latter
ratios are taken from de Grijp et al. (1992).

Strong correlations between the (total or central) radio power, the
widths of narrow emission lines and the luminosities of the narrow
emission lines (a measure of the power of ionizing photons reaching
the NLR) have been known for many years in Seyferts (e.g. Whittle,
1992b, c). Also Seyferts having greater infrared luminosities tend to
display broader narrow emission lines (e.g., Veilleux, 1991); the same
behaviour is observed in generic samples of luminous infrared galaxies
(e.g., Veilleux et al., 1995). Remarkably, Miller et al. (1993) noted
that the the [OIII]$\lambda5007$\AA $\!$ luminosities are more closely
related to the radio core than to extended radio powers in a sample of
low--redshift (z$<$0.5) quasars. All these correlations probably arise
from interactions of ejected radio plasma with the ambient
interstellar medium (e.g., the review by Wilson, 1992).

By outlining the relation between radio core 
and narrow line emissions, which is essentialy driven by S2s,
our study suggests that the energy which powers radio cores 
is fundamentally linked to the supply of energy for the central UV source 
which photoionizes the NLR.
The absence of correlation between radio core power and narrow line
width might be due to small--number statistics.

\subsection{Radio cores and interaction strength.}

In order to describe the interaction strength on our objects, we
consider some parameters defined in the literature: \begin{enumerate}
\item the dimensionless interaction class IAC defined by Dahari (1985)
as an integer which grows with the interaction effect on the galaxies
as seen projected on the sky. The parameter IAC describes both the
degree to which a galaxy looks disturbed and the closeness to a
neighbour (if there is a close companion). Dahari (1985) assigned
IAC=1 to isolated or symmetric galaxies and IAC=6 to highly distorted
galaxies or overlapping objects. Later, many Seyfert galaxies were
assigned IAC-values by Dahari \& De Robertis (1988a). The value of
this IAC parameter is available for 41 objects in common
($N_{ul}$=24). \item the interaction class (hereafter IAC') defined on
a scale 1 to 6 by Whittle (1992) as an indicator of the presence of a
companion (relative size and proximity). Unlike the previous
parameter, this quantity monitors only possible tidal perturbations by
nearby companions and not the degree of morphological disturbance. It
is tabulated by Whittle (1992a) and Nelson \& Whittle (1995) for 48
objects ($N_{ul}$=29) in common. \item the disturbance class DC
defined on a scale of 1 to 6 by Whittle (1992a) as an indicator of a
visible response, either to tidal or to some other perturbation. It is
given by Whittle (1992a) and Nelson \& Whittle (1995) for 43 objects
($N_{ul}$=28). \item the perturbation class PC, a more comprehensive
parameter, which we define as Maximum (IAC, IAC', DC). It is available
for 66 objects ($N_{ul}$=39).
\item the presence of a close companion of a similar redshift (if its
redshift is known). We take this kind of information generally from
Dahari (1994,1995), Dahari \& De Robertis (1988a), MacKenty (1990),
Heisler \& Vader (1994), Rafanelli, Violato \& Baruffolo (1995), and
the catalogue by Lipovetski et al. (1987). This information is
available for 91 objects ($N_{ul}$=51). In general the selection of
probable physical companions of unknown redshift is based on the
amount of magnitude difference and separation between the companion
and the main component. Although in most cases different authors agree
on the presence or absence of a close companion, there is a serious
disagreement in the literature regarding 12 ambiguous cases. We simply
subdivide the 90 galaxies (for which this kind of information is
available into objects without companions (to which we assign a
parameter INT=0), objects with companions (denoted by INT=2), and
ambiguous cases (denoted by INT=1).  \end{enumerate}

Employing the five two--sample tests, we find no significant
difference in the $\Delta(\log P_c)$ distributions between objects
with large ($>3$ or $>2$) and small ($<4$ or $<3$) values of the
parameters IAC, IAC', DC, PC (although objects with large PC--values
of these parameters tend to host slightly more frequently detected
radio cores).

\begin{figure}[h]
\plotone{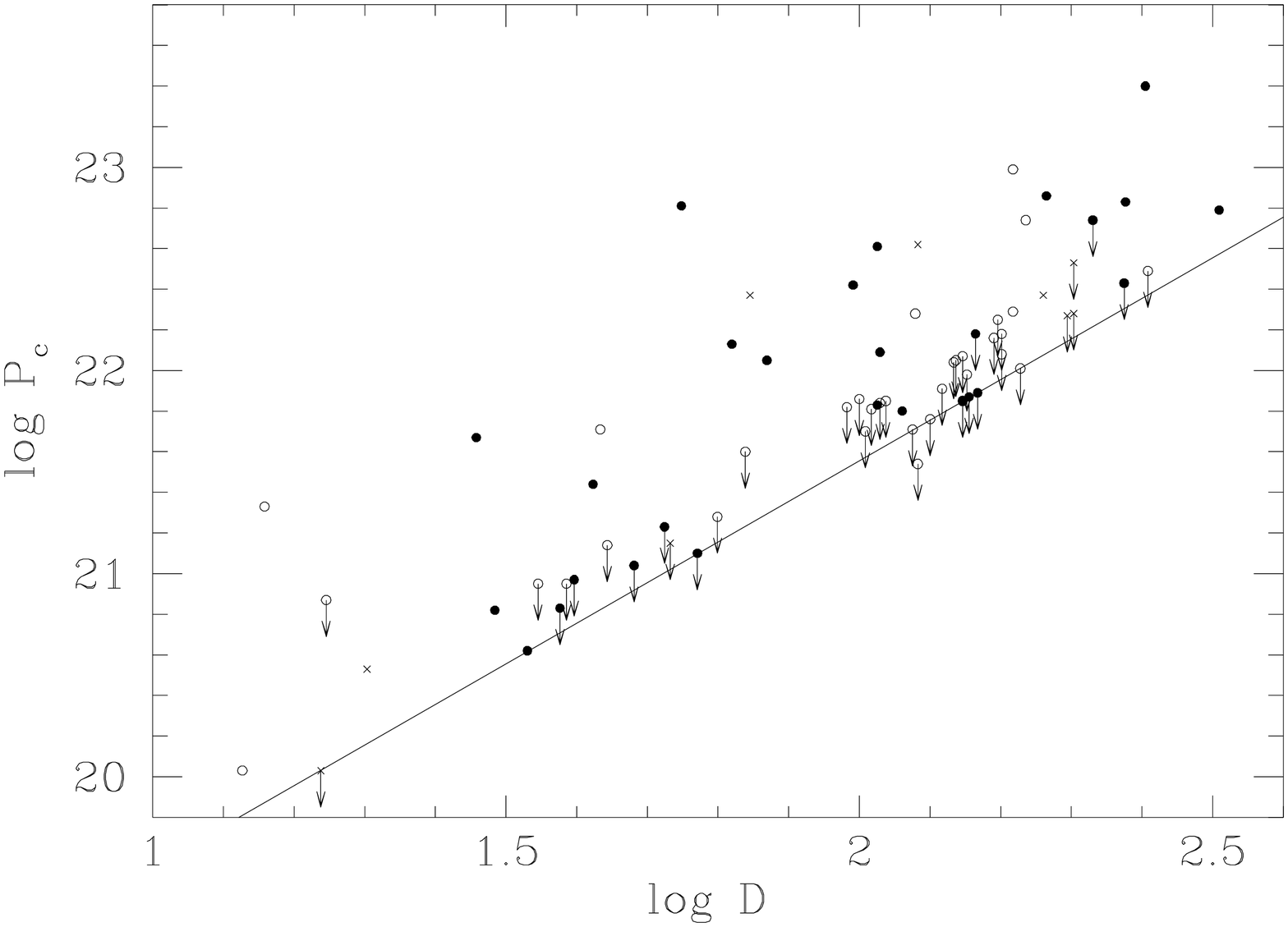}
\caption{The $\log P_c$--$\log D$ plot (as in Figure 2),
but only for non-E/S0 galaxies with some information on the presence
of a nearby companion. $P_c$ is the radio core power (in W/Hz) and $D$
is the distance (in Mpc).  Different symbols denote galaxies without a
companion (open circles), galaxies with a companion (dots), and
uncertain cases (crosses).  The straight line given by eq. (1) in the
text is shown.}
\end{figure}

On the other hand, galaxies with companions (with INT=2) tend to have
a $\Delta(\log P_c)$-- distribution slightly shifted to greater values
than galaxies without companions (with INT=0), with an average
statistical significance of $\sim$94\%. The inclusion of INT=1 objects
in the INT=0 group or in the INT=2 group confirms this result with a
slightly reduced statistical significance. But, if we exclude E/S0
objects, this effect becomes much stronger ($>99$\% significance
level) and amounts to a median difference of a factor of $\sim$3 in
power.  Furthermore, we have checked that this effect is not simply
due to a slight difference in the S2/S1 number ratio between INT=0 and
INT=2. Table 6 presents the most interesting results. Figure 6 shows
the $\log P_c-\log D$ diagram for the non-E/S0 galaxies with INT=0
(open circles), INT=1 (crosses), and INT=2 (dots).

We conclude that radio core activity is favoured by the presence of a
close companion, at least in spiral galaxies.  Our finding appears to
be in line with the results of the VLA radio survey of luminous IRAS
galaxies by Neff \& Hutchings (1992). They noted that in their sample,
which includes also many non-Seyfert galaxies, sources with detected
radio cores (of $<$0.8" size) reside preferentially in systems with a
greater interaction strength, which was established mainly on the
basis of morphological disturbances (Hutchings \& Neff,
1991). Incidentally, interacting Seyfert galaxies were found to
display also enhanced radio emissions from the entire galaxy and from
the central ($\lesssim 20"$) region (e.g., Dahari \& De Robertis,
1988b; Giuricin et al., 1990).

\section{Summary and Conclusions}

We may summarize the principal results of our paper as follows:

1) The compact radio cores of Seyfert galaxies are characterized by a
median power of $\log P_c$= 20.9 (W/Hz) together with a median
core-to-total power ratio of 0.17. They are roughly comparable in
power with the radio cores observed in the nearest (z$<$0.1) objects
of the Bright Quasar Sample, whilst they appear to be typically weaker
than the cores (frequently) detected in early--type galaxies and
stronger than the analogous radio structures (rarely) detected in
non--Seyfert spiral galaxies.

The radio core power correlates with the total radio power, ($P_c
\propto P_T^{1.0\pm0.2}$), and the extended radio power ($P_T-P_c$),
but it appears to be unrelated to the structural characteristics of
extended radio emission (on an arcsec angular scale), which is not
dominated by compact radio cores.

2) Seyfert nuclei hosted in early--type galaxies show a somewhat
stronger radio core emission (typically by a factor of $\sim$2) than
the norm for Seyfert objects.  This appears to reflect the general
strong association of radio--loud objects with early--type
galaxies. Moreover, early--type Seyferts tend to show a slightly
flatter $P_c$--$P_T$ relation ($P_c \propto P_T^{0.8\pm0.2}$) than the
average for the whole sample. Our results indicate that also within
the Seyfert class a connection between radio core activity and galaxy
morphology is present (see, e.g., Wilson \& Colbert, 1995; Fabian \&
Rees, 1995 for recent, different interpretations of this connection).

3) Galaxies with a nearby companion (especially non--early--type ones)
display enhanced radio core emission with respect to objects without
one.  This effect, which is in line with the behaviour of extended
radio emission, lends further support to the idea that Seyfert
activity is stimulated by interactions (e.g. Osterbrok's (1993)
review), especially in early--type spirals (e.g., Monaco et al.,
1994).

4) Seyfert 2 galaxies are confirmed to have more powerful radio cores
(typically by a factor of $\sim$1.5) than Seyfert 1. We have verified
that this is not due to a different proportion of early types or
objects with nearby companions in the two Seyfert classes.

5) Radio core emission is unrelated to optical, near--infrared,
far--infrared emissions and ratios of some prominent narrow emission
lines. In this respect, the behaviour of the radio core emission
appears to be markedly different from that of total and extended radio
emission, which well correlate with the optical and far--infrared
emissions of the host galaxy.

6) Radio core emission shows some relation with prominent (albeit
dissimilar in nature) signatures of AGN activity, such as hard X-ray,
mid--infrared (in the IRAS $\sim12\mu$m and $\sim25\mu$m bands), and
narrow--line ([OIII]$\lambda5007$\AA, $H\beta$) emissions. In this
case, the behaviour of radio cores preserves some similarity with that
of extended radio power, which is known to be related to the same
emissions.
 
The last results (point 6) suggest that Seyfert radio cores are
typically powered by AGNs rather than by radio supernovae, although
the power of radio supernovae (e.g., Weiler et al., 1986; Colina \&
P\'erez--Olea, 1992) is comparable with that of the weakest detected
radio cores in Seyferts.

Our results (points 1) and 2)) emphasize the difference (Sadler et
al., 1995) between Seyfert radio cores and cores of early--type
galaxies. On the other hand, a comparison with the compact radio cores
observed in nearby QSOs is hardly feasible.  Although in many respects
the latter structures do not seem identical to Seyfert cores, the
discrepancies could be entirely accounted for by differences in
spatial resolution and radio survey sensitivity (see, e.g., the end of
\S 2.4 for a discussion of the radio core--optical relation in
Seyferts and QSOs).  In this respect, we deem it interesting that
radio core strength appears to correlate with mid--infrared emission,
since mid--infrared, being an approximate constant fraction of the
bolometric flux, is believed to provide the best indication of the
bolometric luminosity of Seyferts (e.g., Spinoglio \& Malkan, 1989;
Spinoglio et al., 1995).

Therefore, we suggest that the radio core power is likely to be linked
to the bolometric luminosity of Seyfert nuclei, despite the fact that
it represents a negligible fraction of the power output of the central
engine. This link would assimilate Seyfert radio cores to radio--quiet
QSO radio cores, which have recently been claimed to be reliable
tracers of the QSO bolometric power output (Lonsdale et al., 1995),
which in QSOs, unlike Seyferts (precisely Seyferts 2), is typically
dominated by blue and UV radiation.

Besides high--resolution observations at higher frequencies (less
affected by free--free absorption), there is a strong need for radio
measurements on an arcsec/subarsec scale for many Seyfert
objects. These measurements would probe the connection (if any)
between compact radio cores and inner radio structures, for which
limited and inhomogeneous information is now available.

\acknowledgments The authors are grateful for the ASURV software
package kindly provided by E.  D. Feigelson and for a computer code on
correlation analysis kindly given by M.  Akritas. One of the authors
(G. G.) thanks M. Kukula for useful discussions.

This research has made use of the NASA/IPAC Extragalactic Database
(NED), which is operated by the Jet Propulsion Laboratory, Caltech,
under contract with the National Aeronautics and Space
Administration. This work has been partially supported by the Italian
Ministry of University, Scientific and Technological Research (MURST)
and by the Italian Space Agency (ASI).

\onecolumn

\input{./tables/table1}
\input{./tables/table2}
\input{./tables/table3}

\input{./tables/table4}
\input{./tables/table5}
\input{./tables/table6}

\twocolumn

\end{document}

%% file: tables/table1.tex
\begin{table}
\caption{Comparisons between the Radio Core Power of Objects 
of Different Morphologies}
\small
\begin{tabular}{ccrrrrrrr}
\hline
Sample pairs  & Variable   & 
N             & $N_{ul}$   & 
p($G_1$)      & p($G_2$)   & 
p(L)          & p($P_1$)   & 
p($P_2$) \\
\hline
E/S0  &                &31  &15&     &     &     &     &       \\
vs    &$\Delta(\log P)$&    &  &0.075&0.070&0.062&0.069& 0.140 \\
S     &                & 72 &49&     &     &     &     &       \\
\hline
E/S0  &                &31  &15&     &     &     &     &       \\
vs    &$\Delta(\log P)$&    &  &0.049&0.045&0.039&0.045& 0.107 \\
S+S/S0&                & 76 &53&     &     &     &     &       \\
\hline
E/S0  &                &31  &15&     &     &     &     &       \\
 vs   & $\Delta(\log P)$&   &  &0.092&0.085&0.063&0.078& 0.213 \\
S+S/S0+uncl.&          &118 &78&   &     &     &     &         \\
\hline
\end{tabular}
\vspace{-0.5cm}
\end{table}
\normalsize

%% file: tables/table2.tex
\begin{table}
\caption{Correlations between Radio Core Power and Infrared Quantities}
\small
\begin{tabular}{lcrrrrrr}
\hline
Variables    &  Sample   &  
N            &  $N_{ul}$ & 
$N'_{ul}$    &  $\tau$   & $\sigma$     &  $p$      \\
\hline
$\log P_c-\log L_N$            &all& 55& 29& 3&0.025&0.045&$>$0.10 \\
$\log P_c-\log L_{12}$         & S1& 54& 40&19&0.046&0.027&   0.09 \\
$\log P_c-\log L_{12}$         & S2& 88& 47&36&0.056&0.032&   0.08 \\
$\log P_c-\log L_{12}$         &all&144& 89&57&0.053&0.022&   0.02 \\
$\log P_c-\log L_{25}$         & S1& 54& 40& 7&0.008&0.029&$>$0.10 \\
$\log P_c-\log L_{25}$         & S2& 88& 47& 3&0.065&0.033&   0.05 \\
$\log P_c-\log L_{25}$         &all&144& 89&10&0.038&0.023&   0.10 \\
$\log P_c-\log L_{60}$         &all&144& 89& 4&0.002&0.026&$>$0.10 \\
$\log P_c-\log L_{100}$        &all&144& 89&21&0.018&0.025&$>$0.10 \\
\hline					       
$\log P_c-\log (F_{12}/F_{60})$&all&140& 85&53&0.058&0.025&   0.02 \\
$\log P_c-\log (F_{25}/F_{60})$& S1& 51& 37& 4&0.020&0.048&$>$0.10 \\
$\log P_c-\log (F_{25}/F_{60})$& S2& 87& 46& 2&0.079&0.036&   0.03 \\
$\log P_c-\log (F_{25}/F_{60})$&all&140& 85& 6&0.064&0.028&   0.02 \\
\hline
\end{tabular}
\normalsize
\end{table}

%% file: tables/table3.tex
\begin{table}
\caption{Correlation between Radio Core Power and Hard X-Ray Emission}
\small
\begin{tabular}{crrrrrr}
\hline
Variables   &  N          &  
$N_{ul}$    &  $N'_{ul}$  &
$\tau$      &  $\sigma$   &
$p$         \\
\hline
$\log P_c-\log L_x$ & 30 & 17 & 2 & 0.144 & 0.062 & 0.021 \\
\hline
\end{tabular}
\vspace{-0.5cm}
\normalsize
\end{table}

%% file: tables/table4.tex
\begin{table}
\caption{Correlation between Core and Total Radio Powers}
\small
\begin{tabular}{ccrrrrrr}
\hline
Variables     &   Sample   &
N             &   $N_{ul}$ &
$N'_{ul}$     &   $\tau$   & 
$\sigma$      &   $p$     \\  
\hline
$\log P_c-\log P_T$ &  S2& 44 & 20 &  4 & 0.153 & 0.070 & 0.028 \\
$\log P_c-\log P_T$ & all& 92 & 55 & 13 & 0.076 & 0.037 & 0.041 \\
\hline
\end{tabular}
\normalsize
\end{table}

%% file: tables/table5.tex
\begin{table}
\caption{Correlation between Radio Core Powers and Emission Lines}
\small
\begin{tabular}{ccrrrrrr}
\hline
Variables       &    Sample    &
N               &    $N_{ul}$  &
$N'_{ul}$       &    $\tau$    &
$\sigma$        &    p         \\
\hline
$\log P_c-\log L_{[OIII]}$& all&130 & 80 &  0 & 0.040 & 0.022 & 0.069 \\
$\log P_c-\log L_{H\beta}$&  S2& 79 & 41 & 21 & 0.083 & 0.027 & 0.003 \\
$\log P_c-\log L_{H\beta}$& all& 99 & 56 & 21 & 0.073 & 0.024 & 0.003 \\
\hline
\end{tabular}
\normalsize
\end{table}

%% file: tables/table6.tex
\begin{table}
\caption{Comparison between the Radio Core Power of Objects with and 
without Close Companions}
\small
\begin{tabular}{ccrrrrrrr}
\hline
Sample Pairs    &    Variable   &
N               &    $N_{ul}$   &
p($G_1$)        &    p($G_2$)   &
p(L)            &    p($P_1$)   &
p($P_2$)        \\
\hline
INT=0   &                &47&32&     &     &     &     &      \\
 vs     &$\Delta(\log P)$&  &  &0.083&0.081&0.038&0.062&0.056 \\
INT=2   &                &32&13&     &     &     &     &      \\     
\hline
\multicolumn{9}{c}{Without E/S0 galaxies}\\
\hline
INT=0   &                &34&27&     &     &     &     &      \\  
  vs    &$\Delta(\log P)$&  &  &0.009&0.008&0.004&0.005&0.005 \\
 INT=2  &                &27&11&     &     &     &     &      \\  
\hline
\end{tabular}
\normalsize
\end{table}

%% file: seyfert.bbl
\begin{references}

\reference{} Adams, T. F., 1977, \apjs, 33, 19.

\reference{} Antonucci, R. R. J., \araa, 31, 473.

\reference{} Antonucci, R. R. J. \& Barvainis, R., 1988, \apj, 332,
L13.

\reference{} Akritas, M. G. \& Siebert, J., 1995, \mnras, in press.

\reference{} Awaki, H., Koyama, K., Inoue, H., Halpern, J., 1991,
\pasj, 43, 195.

\reference{} Baker, J. C., Hunstead, R. W. \& Brinkmann, W., 1995,
\mnras, 277, 553.

\reference{} Bicay, M. D., Kojoian, G., Seal, J., Dickinson, D. F.,
Malkan, M. A., 1995, \apjs, 98, 369.

\reference{} Boller, Th. et al., 1992, \aap, 261, 57.

\reference{} Burns, J. O. et al., 1987, \aj, 94, 587.

\reference{} Cappi, M., Mihara, T., Matsuoka, M., Brinkmann, W.,
 Prieto, M. A., Palumbo, G. G. C., 1995, ApJ, in press.

\reference{} Colina, L. \& P\'erez--Olea, D., 1993, \mnras, 259, 709.

\reference{} Colina, L. \& P\'erez--Olea, D., 1993, in "The Nearest
Active Galaxies", ed. J. E. Beckman (Dordrecht: Kluwer Academic Pub.),
99.

\reference{} Comastri, A. et al., 1992, \apj, 384, 62.

\reference{} Condon, J. J., 1987, \apjs, 65, 485.

\reference{} Condon, J. J. \& Broderick, J. J., 1988, \aj, 96, 30.

\reference{} Condon, J. J., Helou, G., Sanders, D. B., Soifer, B. T.,
1990, \apjs, 73, 359.

\reference{} Cram, L. E., North, A. \& Savage, A., 1992, \mnras, 257,
602.

\reference{} Dahari, O., 1984, \aj, 72, 730.

\reference{} Dahari, O., 1985, \apjs, 57, 643.

\reference{} Dahari, O. \& De Robertis, 1988a, \apjs, 67, 249.

\reference{} Dahari, O. \& De Robertis, 1988b, \apj, 331, 727.

\reference{} Danese, L., Zitelli, V., Granato, G. L., Wade, R.,De
Zotti, G.  \& Mandolesi, N., 1992, \apj, 399, 38.

\reference{} De Grijp, M. H. K., Keel, W. C., Miley, G. K., Goudfrooj,
P. \& Lub, J., 1992, \aaps, 96, 389.

\reference{} Della Ceca, R.  et al., 1994, \apj, 430, 533.

\reference{} de Vaucouleurs, G., de Vaucouleurs, A., Corwin, H. G.,
Buta, R. J., Paturel, G. \& Fouqu\'e, 1991, "Third Reference Catalog
of Bright galaxies" (New York: Springer Verlag) (RC3).

\reference{} Durret, F., 1989, \aaps, 81, 253.

\reference{} Edelson, R. A., 1987, \apj, 313, 651.

\reference{} Edelson, R. \& Malkan, M. A., 1987, \apj, 321, 233.

\reference{} Fabbiano, G., Kim, D.-W. \& Trinchieri, G., 1992, \apjs,
80, 531.

\reference{} Fabbiano, G., Miller, L., Trinchieri, G., Longair, M. \&
Elvis, M., \apj, 277, 115.

\reference{} Fabian, A. C. \& Rees, M. J., 1995, \mnras, 277, L55.

\reference{} Feigelson, E. D., 1992, in "Statistical Challenges in
Modern Astronomy" (Springer Verlag), p. 221.

\reference{} Feigelson, E. D. \& Nelson, P. I., 1985, \apj, 293, 192.

\reference{} Fullmer, L. \& Lonsdale, C. J., 1989, "Cataloged Galaxies
and Quasars Observed in the IRAS survey", 2nd version, California
Institute of Technology, Pasadena, California.

\reference{} Gezari, D. Y., Schmitz, M., Pitts, P. S. \& Mead, J. M.,
1993, "Catalog of Infrared Observations", NASA Reference Publication
No. 1294.

\reference{} Gioia, I. M., Feigelson, E. D., Maccacaro, T., Schild,
R., Zamorani, G., 1983, \apj, 271, 524.

\reference{} Giuricin, G., Limboz Tektunali, F., Monaco, P.,
Mardirossian, F., Mezzetti, M., 1995a, \apj, 450, 41.

\reference{} Giuricin, G., Mardirossian, F., Mezzetti, M., 1995b,
\apj, 446, 530.

\reference{} Giuricin, G., Mardirossian, F., Mezzetti, M. \& Bertotti,
G., 1990, \apjs, 72, 551.

\reference{} Ghosh, K. K. \& Soundararajaperumal, S., 1992, \mnras,
259, 175.

\reference{} Granato, G. L., Zitelli, V., Bonoli, F., Danese, L.,
Bonoli, C., Delpino, F., 1993, \apjs, 89, 35.

\reference{} Grandi, P., Tagliaferri, G., Giommi, P., Barr, P.,
Palumbo, G. G. C., 1992, \apjs, 82, 93.

\reference{} Green, P. J., Anderson, S. F. \& Ward, 1992, \mnras, 254,
30.

\reference{} Gregorini, L. et al., 1994, \aaps, 106, 1.

\reference{} Gregory, P. C. \& Condon, J. J., 1991, \apjs, 75, 1011.

\reference{} Grossan, B. A., 1992, PhD. thesis, Massachussetts
Institute of Technology.

\reference{} Isobe, T., Feigelson, E. D. \& Nelson, P. J., 1986, \apj,
306, 490.

\reference{} Halpern, J. P., 1982, Ph.D. thesis, Harvard University.

\reference{} Harnett, J. I., 1987, \mnras, 257, 602.

\reference{} Heisler, C. A. \& Vader, J. P., 1994, \aj, 107, 35.

\reference{} Helou, G., Khan, I. R., Malek, L. \& Boehmer, L., 1988,
\apjs, 68, 151.

\reference{} Hill, G. J., Becklin, E. E. \& Wynn--Williams, 1988,
\apj, 330, 737.

\reference{} Hutchings, J. B. \& Neff, S. G., 1991, \aj, 101, 434.

\reference{} Kellermann, K., Sramek, R., Schmidt, M., Shaffer,
D. B. \& Green, R., 1989, \aj, 98, 1195.

\reference{} Kendall, M. \& Stuart, A., 1977, The Advanced Theory of
Statistics (London: Griffin \& Co.).

\reference{} Kirhakos, S. D. \& Steiner, J. E., 1990, \aj, 99, 1435.

\reference{} Klein, U., Weiland, H. \& Brinks, E., 1991, \aap, 246,
323.

\reference{} Kotilainen, J. K. \& Prieto, M. A., 1995, \aap, 295, 646.

\reference{} Kotilainen, J. K. \& Ward, M. J., 1994, \mnras, 266, 953.

\reference{} Kotilainen, J. K., Ward, M. J., Boisson, C., DePoy,
D. L., Bryant, L. R., Smith, M. G., 1992a, \mnras, 256, 125.

\reference{} Kotilainen, J. K., Ward, M. J., Boisson, C., DePoy,
D. L., Bryant, L. R., Smith, M. G., 1992b, \mnras, 256, 149.

\reference{} Kotilainen, J. K., Ward, M. J. \& Williger, G. M., 1993,
\mnras, 263, 655.

\reference{} Koyama, K. et al., 1989, \pasj, 41, 731.

\reference{} Kormendy, J. \& Richstone, D., 1995, \araa, 33, 581.

\reference{} Kruper, J.S., Urry, C. M. \& Canizares, 1990, \apjs, 74,
347.

\reference{} Kukula, M. A., Pedlar, A., Baum, S. \& O'Dea, C. P.,
1995, \mnras, 276, 1262.

\reference{} Lauberts, A. \& Valentijn, E. A., 1989, "The Surface
Photometry Catalogue of the ESO--Uppsala Galaxies" (Garching bei
Munchen: European Southern Observatory).

\reference{} Lawrence, A., 1987, \pasp, 99, 309.

\reference{} La Valley, M., Isobe, T. \& Feigelson, E. D., 1992,
\baas, 24, 839.

\reference{} Lipovetski, V. A., Neizvestny, S. J. \& Neizvestnaya,
 O. M., 1987, "A Catalogue of Seyfert Galaxies", Communications of the
 Special Astrophysical Observatory, 55.

\reference{} Lo, K. Y., 1994, in "The Nuclei of Normal Galaxies",
R. Genzel \& A. I. Harris (Dordrecht: Kluwer Academic Pub.), p. 395.

\reference{} Lonsdale, C. J., Smith, H. E. \& Lonsdale, C. J., 1993,
\apj, 405, L9.

\reference{} Lonsdale, C. J., Smith, H. E. \& Lonsdale, C. J., 1995,
\apj, 438, 632.

\reference{} MacKenty, J. W., 1990, \apjs, 72, 231.

\reference{} Maiolino, R., Ruiz, M., Rieke, G. H., Keller, L. D.,
1995, \apj, 446, 561.

\reference{} Malaguti, G., Bassani, L. \& Caroli, E., 1994, \apjs, 94,
517.

\reference{} Mazzarella, J. M., Bothun, G. D. \& Boroson, T. A., 1991,
\aj, 101, 2034.

\reference{} McAlary, W. \& Rieke, G. H., 1988, \apj, 333, 1.

\reference{} McLeod, K. K. \& Rieke, G. H., 1995, \apj, 441, 96.

\reference{} Miller, P., Rawlings, S. \& Saunders, R., 1993, \mnras,
263, 425.

\reference{} Monaco, P., Giuricin, G., Mardirossian, F. \& Mezzetti,
M., 1994, \apj, 436, 576.

\reference{} Nandra, K. \& Pounds, K. A., 1994, \mnras, 268, 405.

\reference{} Neff, S. G. \& Hutchings, J. B., 1992, \aj, 103, 1746.

\reference{} Nelson, C. H. \& Whittle, M., 1995, \apjs, 99, 67.

\reference{} Netzer, H. \& Laor, A., 1993, \apj, 404, L51.

\reference{} Norris, R. P., 1988, \mnras, 230, 345.

\reference{} Norris, R. P., Allen, D. A., Sramek, R. A., Kesteven,
  M. J.  \& Troup, E. R., 1990, \apj, 359, 291.

\reference{} Osterbrock, D. E., 1993, \apj, 404, 551.

\reference{} Osterbrock, D. E. \& Martel, A., 1993, \apj, 414, 552.

\reference{} Rafanelli, P., Violato, M. \& Baruffolo, A., 1995, \aj,
109, 1546.

\reference{} Rao, A. R., Singh, K. P. \& Vahia, M. N., 1992, \mnras,
255, 197.

\reference{} Roberts, M. S. \& Haynes, M. P., 1994, \araa, 32, 115.

\reference{} Rodriguez--Espinosa, J. M., Rudy, R. J. \& Jones, B.,
1987, 312, 555.

\reference{} Roy, A. L., Norris, R. P., Kesteven, M. J., Reynolds,
J. E.  \& Troup, E. R., 1996, in preparation.

\reference{} Roy, A. L., Norris, R. P., Kesteven, M. J., Troup, E. R.
\& Reynolds, J. E., 1994, \apj, 432, 496.

\reference{} Rudy, R. J. \& Rodriguez--Espinosa, J. M., 1985, \apj,
298, 614.

\reference{} Rush, B., Malkan, M. A. \& Spinoglio, 1993, L., 1993,
\apjs, 89, 1.

\reference{} Sadler, E. M., Slee, O. B., Reynolds, J. E. \& Roy,
A. L., 1995, \mnras, 276, 1373.

\reference{} Sanders, D. B., Phinney, E. S., Neugebauer, G., Soifer,
B. T., Matthews, K., 1989, \apj, 347, 29.

\reference{} Saxton, R. D. et al., 1993, \mnras, 262, 63.

\reference{} Siegel, S., 1956, Nonparametric Statistics for
Behavioural Sciences (New--York: McGraw--Hill), 68.

\reference{} Simien, F. \& de Vaucouleurs, G., 1986, \apj, 302, 564.

\reference{} Singh, K. P., Rao, A. R. \& Vahia, M. N., 1991, \aap,
248, 37.

\reference{} Slee, O. B., Sadler, E. M., Reynolds, J. E. \& Ekers,
R. D., 1994, \mnras, 269, 928.

\reference{} Smith, E. P., Heckman, T. M., Bothun, G. D., Romanishin,
W. \& Balick, B., 1986, \apj, 306, 64.

\reference{} Sopp, H. M. \& Alexander, P., 1991, \mnras, 251, 14p.

\reference{} Spinoglio, L. \& Malkan, M. A., \apj, 342, 83.

\reference{} Spinoglio, L., Malkan, M. A., Rush, B., Carrasco, L. \&
Recillas-Cruz, E., 1995, \apj, 453, 616.

\reference{} Telesco, C. M., 1988, \araa, 26, 343.

\reference{} Tovmassian, H. M. et al., 1984, \aaps, 58, 317.

\reference{} Tully, R. B., 1988, Nearby Galaxies Catalog (Cambridge:
Cambridge Univ. Press).

\reference{} Tully, R. B. \& Shaya, E. J., 1984, \apj, 281, 31.

\reference{} Turner, T. J. \& Pounds, K. A., 1989, \mnras, 240, 833.

\reference{} Ulvestad, J. S., 1986, \apj, 310, 136.

\reference{} Ulvestad \& Wilson, 1984a, \apj, 278, 544.

\reference{} Ulvestad \& Wilson, 1984b, \apj, 285, 439.

\reference{} Ulvestad, J. S. \& Wilson, A. S., 1989, \apj, 343, 659.

\reference{} Unger, S. W., Lawrence, A., Wilson, A. S., Elvis, M. \&
Wright, A. E., 1987, \mnras, 228, 521.

\reference{} Unger, S. W., Pedlar, A., Booler, R. V. \& Harrison,
 B. A., 1986, \mnras, 219, 387.

\reference{} Unger, S. W. et al., 1989, \mnras, 236, 425.

\reference{} Vader, P. J., Frogel, J. A., Terndrup, D. M. \& Heisler,
C. A., 1993, \aj, 106, 1743.

\reference{} van Driel, W., van den Broek, A. C. \& de Jong, T., 1991,
\aaps, 90, 55.

\reference{} Veilleux, S., 1991, \apjs, 75, 383.

\reference{} Veilleux, S., Kim, D.-C., Sanders, D. B., Mazzarella,
J. M. \& Soifer, B. T., \apjs, 98, 171.

\reference{} Walter, R. \& Fink, H. H., 1993, \aap, 274, 105.

\reference{} Ward, M. J., Done, C., Fabian, A. C., Tennant, A. F. \&
Shafer, R. A., 1988, \apj, 324, 767.

\reference{} Weiler, K. W., Sramek, R. A., Panagia, N., van der Hulst,
J.  M., Salvati, M., 1986, \apj, 301, 790.

\reference{} White, R. L. \& Becker, R. H., 1992, \apjs, 79, 331.

\reference{} Whittle, M., 1992a, \apjs, 79, 49.

\reference{} Whittle, M., 1992b, \apj, 387, 109.

\reference{} Whittle, M., 1992c, \apj, 387, 121.

\reference{} Williams, O. R. et al., 1992, \apj, 389, 157.

\reference{} Wilkes, B. J. \& Elvis, M., 1987, \apj, 323, 243.

\reference{} Wilkes, B., Tananbaum, H., Worrall, D. M. et al., 1994,
\apjs, 92, 53.

\reference{} Wilson, A. S., 1992, in " Physics of Active Galactic
Nuclei", eds. W. J. Duschl \& S. J. Wagner (Springer Verlag:
Heidelberg), 307.

\reference{} Wilson, A. S. \& Colbert, E. J. M., 1995, \apj, 438, 62.

\reference{} Wilson, A. S. \& Meurs, E. J. A., 1982, \aaps, 50, 217.

\reference{} Wilson, A. S. \& Tsvetanov, Z. I., 1994, \aj, 107, 1227.

\reference{} Worrall, D. M., Giommi, P., Tananbaum, H. \& Zamorani,
G., 1987, \apj, 313, 596.

\reference{} Wright, A. E., 1974, \mnras, 167, 273.

\reference{} Zitelli, V., Granato, G. L., Mandolesi, N., Wade, R. \&
Danese, L., 1993, \apjs, 84, 185.

\end{references}
